\documentclass[a4paper,fleqn]{cas-dc}

\usepackage[numbers]{natbib}
\definecolor{cream}{RGB}{222,217,201}

\newcommand{\n}{\tilde{n}}
\newcommand{\F}{\mathcal{F}}
\newcommand{\R}{\mathbf{R}}
\newcommand{\M}{\mathbf{M}}

\renewcommand{\S}{\mathbf{S}}

\renewcommand{\r}{\mathbf{r}}

\begin{document}
\let\WriteBookmarks\relax
\def\floatpagepagefraction{1}
\def\textpagefraction{.001}
\let\printorcid\relax

\shorttitle{}

\shortauthors{}

\title [mode = title]{Characterization of phospholipid-cholesterol bilayers as self-assembled amphiphile block polymers that contain headgroups}




\author[1,2]{Xiaoyuan Wang}%[<options>]
\credit{Data curation, Software, Formal analysis, Validation, Visualization, Writing-original draft, Writing-review and editing}
%\fnmark[1]% Footnote of the first author

\author[3]{Fredric S. Cohen}
\credit{Validation, Formal analysis, Funding acquisition, Methodology, Writing-original draft, Writing-review and editing}
\author[4]{Shixin Xu}
\ead{shixin.xu@dukekunshan.edu.cn}
\credit{Formal analysis, Funding acquisition, Methodology, Supervision, Resources, Validation, Writing-review and editing}
\cormark[1]% Corresponding author indication
\author[5]{Yongqiang Cai}
\ead{caiyq.math@bnu.edu.cn}% Email id of the first author
\cormark[1]% Corresponding author indication
%\ead[url]{orcid.org/0000-0002-2666-0539}% URL of the first author
\credit{Software, Validation, Formal analysis, Funding acquisition, Methodology, Supervision, Writing-review and editing}


\affiliation[1]{organization={LSEC and NCMIS, Institute of Computational Mathematics and Scientific/Engineering Computing (ICMSEC), Academy of Mathematics and Systems Science (AMSS), Chinese Academy of Sciences},
            %addressline={},
            city={Beijing},
%          citysep={}, % Uncomment if no comma needed between city and postcode
           %postcode={},
            country={ China}}

% \author[2]{}%[]

% % Footnote of the second author

% % Email id of the second author

% % URL of the second author

% % Credit authorship

% % Address/affiliation
\affiliation[2]{organization={School of Mathematics and Statistics},
            city={Wuhan},
%          citysep={}, % Uncomment if no comma needed between city and postcode
            %postcode={},
            %state={},
            country={China}}

\affiliation[3]{organization={Department of Physiology and Biophysics, Rush University Medical Center},
            city={Chicago},
%          citysep={}, % Uncomment if no comma needed between city and postcode
            %postcode={},
            state={Illinois},
            country={USA}}
\affiliation[4]{organization={Zu Chongzhi Center, Duke Kunshan University},
             addressline={8 Duke Ave},
            city={Kunshan},
%          citysep={}, % Uncomment if no comma needed between city and postcode
            %postcode={},
            state={Jiangsu},
            country={China}}
\affiliation[5]{organization={School of Mathematical Sciences, Laboratory of Mathematics and Complex Systems, MOE, Beijing Normal University},
% \affiliation[5]{organization={School of Mathematical Sciences, Laboratory of Mathematics and Complex Systems, MOE, Beijing Normal University},
             % addressline={8 Duke Ave},
            city={Beijing},
%          citysep={}, % Uncomment if no comma needed between city and postcode
            postcode={},
            country={China}}

% % Corresponding author text
\cortext[1]{Corresponding author}
% \cortext[2]{Corresponding author}


\begin{abstract}
Cholesterol is known to modulate the structure and function of biological membranes. In this study, we use self-consistent field theory (SCFT) to investigate phospholipid/cholesterol bilayer membranes modeled with two types of diblock copolymers. These copolymer-based bilayers serve as biomimetic platforms with applications in areas such as drug delivery. Our simulations identify a minimum free energy configuration characterized by phospholipid tails tilted relative to the membrane normal. The model quantitatively captures the well-known area condensation effect as cholesterol concentration increases, along with membrane thickening and reduced tilt angle. Thermodynamically, we observe a linear dependence between cholesterol’s chemical potential and its concentration within the 37–50\% range, consistent with experimental results. Additionally, we analyze the effects of block copolymer length and headgroup interactions on bilayer structure. Interactions between phospholipid headgroups and the solvent emerge as the most influential. This work provides a theoretical framework for understanding cholesterol’s regulatory role in membrane structure and mechanics.
\nocite{}%% Remove this line from your manuscript.
\end{abstract}

%Use if graphical abstract is present
%\includegraphics{figures/graphical abstracts.jpg}
%\item A model characterizing the phospholipid-cholesterol bilayer membranes is presented.
%\item Headgroup interactions are crucial in phospholipid–cholesterol bilayer membrane.
%\item Polymer self-assembled bilayer membranes replicate key cholesterol effects on bilayers.
%\item Polymer self-assembly enables the fabrication of biomimetic lipid membranes.

%\nocite{*}

% Keywords
\maketitle
% Main text
\section{Introduction}

Biological cells are rich in membranes. In addition to its plasma membrane, the intracellular space is packed with membranes. Membranes are not merely static barriers that separate two aqueous spaces, but dynamic structures that pinch off and fuse with other membranes to perform a plethora of functions, such as exocytosis and intracellular trafficking \cite{phillips2012physical}.  Lipids, arranged in a bilayer structure, constitute about $99\%$ of membrane molecules on a mole basis. Proteins, much more massive than lipids, account for about $50\%$ of a membrane’s mass and drive biological functions. Artificial vesicles composed of lipids, sometimes with reconstituted proteins, known as liposomes, have often been used to study membrane deformability, fission, and fusion. Less frequently, vesicles composed of lipids and amphiphilic block copolymers have also been used to mimic cellular membranes \cite{kita2005block,palivan2016bioinspired,brodszkij2024advances}.

The overwhelming majority of cholesterol in the human body resides in cell membranes. This lipid affects membrane structure, such as thickness, and the precise orientation of other membrane lipids. The concentration of cholesterol in membranes varies widely according to the membrane. In plasma membranes, cholesterol is the most abundant lipid with concentrations in the $35$ mol$\%$ range. In the endoplasmic reticulum membranes, cholesterol is dilute, in the range of only $4$ mol $\%$ \cite{ray1969lipid,lange1989plasma}.

Theoretical models and methods have been developed to understand how cholesterol affects bilayer membrane structures. These include molecular dynamics simulations \cite{berkowitz2009detailed,toppozini2014structure,grouleff2015influence,ermilova2019cholesterol,chng2022modulation}, dissipative particle dynamics \cite{groot2001mesoscopic,de2010molecular}, and mean-field theory \cite{elliott2005phase,khelashvili2005self,leermakers2007interaction}. Some investigations have focused on the hydrophobic portion of cholesterol, often simplifying it as a headless, rigid rod \cite{khelashvili2005self,huang1999microscopic,wang2023mimicking}. For instance, in a lattice model, cholesterol is represented as a rigid entity. Our group used self-consistent field theory and modeled cholesterol as headless. We found that, depending on conditions, the bilayer membrane can potentially exist in three distinct phases.

Bilayer membranes have not only been experimentally modeled with the use of lipids but with block copolymers as well \cite{ruysschaert2005hybrid, olubummo2012controlling, pippa2013peo, pippa2016chimeric, ozawa2023bilayer, nishimura2019substrate, spanjers2022assembly, de2021evaluation,ulbricht2006advanced}. For instance, hybrid vesicles, consisting of dipalmitoylphosphatidylcholine (DPPC) and triblock copolymers, have had the length of the blocks systematically altered, and impacts determined \cite{dao2017mixing}. In addition, some of these hybrid vesicles are very stable \cite {spanjers2022assembly}, allowing them to be used as artificial organelles \cite{messager2014novel,najer2013polymer}, as targeted drug delivery systems \cite{pawar2013functionalized}, cell mimetics \cite{marguet2013multicompartmentalized}, and nanoreactors \cite{palivan2012protein}.

Our ultimate goal is to utilize two distinct diblock copolymers to model phospholipid/cholesterol systems. In the present study, we examined the impact of increasing cholesterol concentration on bilayer membranes composed of a single phospholipid, such as DPPC, and cholesterol.

The remainder of this paper is organized as follows: Section \ref{SEC:M_M} introduces the self-consistent field theory used to simulate phospholipid/cholesterol bilayers. Section \ref{SCE:R_D} presents results and a discussion that cover parameter configurations, potential phases, the effect of the content of a rod-coil block copolymer (a mimic for cholesterol) on bilayer structure, density distributions, and chemical potentials. Additionally, the influence of geometric asymmetry parameters and interaction strengths on bilayer behavior is discussed. In section \ref{SCE:Conclusion}, we present conclusions and propose some prospective future improvements.

\section{Model and method}
\label{SEC:M_M}
\subsection{Basic model}

Cholesterol consists of a small headgroup, a rigid hydrophobic portion of four fused rigid rings, and a short hydrophobic tail. To simplify models, some theoretical studies, including one of our own, have omitted the headgroup of cholesterol \cite{khelashvili2005self,huang1999microscopic, wang2023mimicking}; However, it is well known that the headgroup is critical for many of cholesterol’s effects on bilayers, perhaps most famously for its contribution to the condensation effect: the area of a cholesterol/saturated PC monolayer or bilayer is less than the sum of the areas occupied by each of the two lipids alone. Block copolymers exhibit rich self-assembly behavior\cite{Jenekhe1998,nunes2020thinking,schacher2009self}, with the ability to form bilayer structures in hydrophilic solvents\cite{cai2017liquid}. It has yet to be determined whether bilayer membranes formed from two different diblock copolymers can effectively describe a phospholipid/cholesterol system, particularly key characteristics such as thermodynamic properties and membrane structure. We analyzed a self-assembled system using a self-consistent field theory (SCFT). The bilayer's free energy was calculated for a canonical ensemble by integrating Flory-Huggins interactions, Maier-Saupe interactions, and stretching energies of flexible polymers.

\begin{figure}[htbp!]
\centering
\includegraphics[width= 1\linewidth]{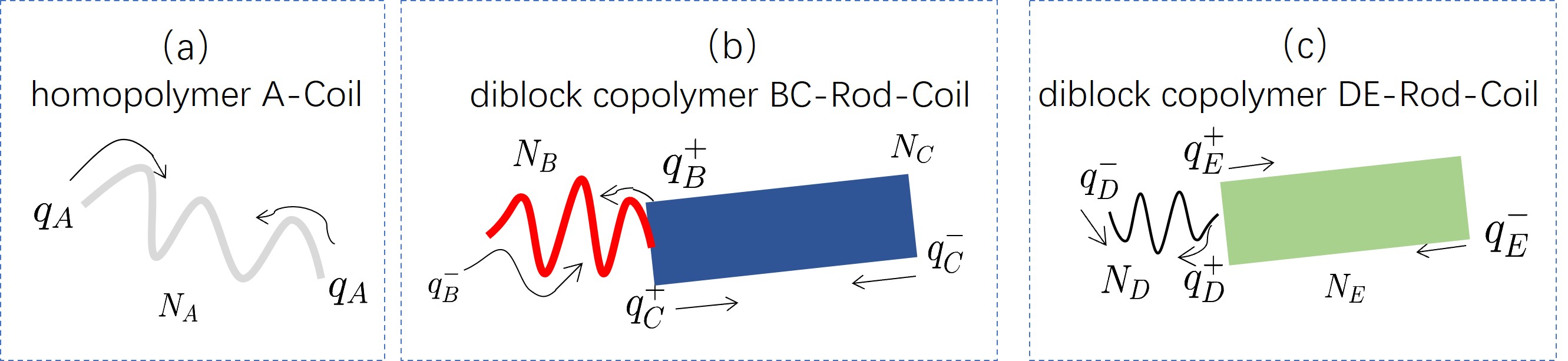}
\caption{Schematic diagrams of polymer structures:
(a) Homopolymer A-Coil with degree of polymerization $N_A$.
(b) Diblock copolymer BC-Rod-Coil with total degree of polymerization $N_B + N_C$.
(c) Diblock copolymer DE-Rod-Coil with total degree of polymerization $N_D + N_E$. The arrows indicate the directions that we used to solve the propagators.
}
\label{fig:ABCDE1}
\end{figure}

The system that we are examining includes three polymers, as shown in Fig.~\ref{fig:ABCDE1}. Water was modeled as a flexible homopolymer, denoted A. The phospholipid was modeled as a BC diblock copolymer, where the headgroup was represented by a flexible polymer, B, and the acyl chains were simulated by a rigid polymer, C. Cholesterol was represented as a diblock copolymer with its headgroup described as flexible, D, and its hydrophobic portion was rigid, E. The BC and DE diblock copolymers were preassembled into bilayer membranes that were surrounded, on both sides, by the A-homopolymer.

We denote the system's volume by $V$. The number of A, BC, and DE polymers is given by $n_{A} \equiv \n_{A} n$, $n_{BC} \equiv n = \tilde n_{BC} n$ (\emph{i.e.}, $\n_{B} = 1$), and $n_{DE} \equiv \tilde n_{DE} n$, respectively. Similarly, the degrees of polymerization are given by $N_{A}$, $ N_{BC}\equiv N$, and $N_{DE}N$, respectively. The volume fractions of the A homopolymer, B-Coil and C-Rod of the BC diblock copolymers, and D-Coil and E-Rod of the DE diblock copolymers, are $f_A=N_A/N$, $f_B=N_B/N$, $f_{C}=N_{C}/N = 1 - N_B/N$, and $f_D = N_D/N$, $f_E = N_E/N$, respectively. To simplify the model, all monomers share the same density $\rho_0$, which is defined as the number of monomers per unit volume \cite{wu2016variational}.
Thus, we have
$
    \frac{\rho_0V}{N}
    =
    n_{BC} + n_{A}f_A + n_{DE} f_{DE}
    =
    n(1+\n_{A}f_{A}+\n_{DE}f_{DE})
    =:n\tilde{\rho}
$, where we refer to $\tilde \rho$ as the effective volume fraction.

The flexible coils A, B, and D are distinct, as are the rigid C and E rods. The statistical segment lengths of the A, B, C, D, and E blocks are represented as $\tilde{a}$, $\tilde{b}$, $\tilde{c}$, $\tilde{d}$, and $\tilde{e}$, respectively. The geometrical asymmetry parameters for the rods are expressed as $\beta_C = \tilde{c}N / R_g$ \cite{song2011phaseJPCB} and $\beta_E = \tilde{e}N / R_g$,  where $R_g = \sqrt{N \tilde{b}^2 / 6}$.

The system's interactions consist of isotropic Flory-Huggins interactions and anisotropic Maier-Saupe interactions. The isotropic interaction potentials \cite{flory1953principles,ohta1986equilibrium} between  molecules are denoted as $ H_{FH}=\rho_0\int_V [
    \sum\limits_{i \neq j}
        \chi_{ij} \hat{\phi}_i(\r)\hat{\phi}_{j}(\r)
       ]d\r$, $\chi_{ij}(i,j=A,B,C,D,E, i \neq j)$ is the Flory-Huggins parameter.
The anisotropic Maier-Saupe interaction potential between molecules
\cite{maier1958einfache,li2014phase} is $H_{MS}=-\frac{\eta \rho_0}{2}\int_V \hat{S}(\r):\hat{S}(\r)d\r$, where $\eta$ is the Maier-Saupe parameter that describes the strength of orientation.

The orientational order parameter is defined as $\hat{S}= \langle \mathbf{u}\mathbf{u}-\frac{\mathbf{I}}{3}\rangle$, where $\mathbf{u}$ is a unit vector, and $\langle \cdot \rangle$ denotes the average over the molecular distribution. The Maier-Saupe potential facilitates a parallel alignment of the rigid rods. The stretching energy of the flexible coils is expressed as
$H_{stretching} = \frac{3}{2b^2}
            \sum\limits_{k}
            \int_{0}^{N_k} \big|
            \frac{\textrm{d} \R_{k^{'}}(s)}{\textrm{d}s} \big|^2 \textrm{d}s
$, where the summation over $k$ accounts for each coil block. Within the SCFT framework formulated in the canonical ensemble \cite{wu2016variational}, the Helmholtz free energy per chain of the system is expressed as:
\begin{align}\label{energy:Hamiltonian1}
\frac{N\F}{V\rho_0 k_BT}
=&
      \frac{1}{V}\int_V
      d\mathbf{r}\big[\sum\limits_{i \neq j}\chi_{ij}N \phi_i(\mathbf{r})\phi_j(\mathbf{r})\notag\\
       &- \sum\limits_{i }\omega_i(\mathbf{r})\phi_i(\mathbf{r})
       + \M(\mathbf{r}):\S(\mathbf{r})\notag\\
    &-\frac{\eta N}{2}\S(\mathbf{r}):\S(\mathbf{r})
       -\xi(\mathbf{r})(\sum\limits_{i }\phi_i(\mathbf{r})-1)\big]\notag\\
      &- \frac{1}{\tilde{\rho}} \log (Q_{A}^{\n_A} Q_{BC}^{\n_{BC}} Q_{DE}^{\n_{DE}}),
\end{align}
where $\phi_i(\r)$ and $\omega_i(\r)$ represent the local density concentrations and the mean fields of the $i$-type monomers ($i=A,B,C,D,E$), respectively. $\S(\r) =\S_C(\r) + \S_E(\r)$ denotes the local orientational density distribution of the rigid blocks, and $\M(\r)$ refers to the mean fields associated with these blocks. $\xi(\r)$ is the Lagrange multiplier that enforces incompressibility of the system. The terms $Q_{A}[\omega_A]$, $Q_{BC}[\omega_B,\omega_C,\M]$, and $Q_{DE}[\omega_D,\omega_E,\M]$ correspond to the partition function contributions of the polymers' single-chain behavior in the presence of their respective mean fields. Further details on the model and the numerical methods used for its solution are provided in the Supporting Materials.

\section{Results and discussion}
\label{SCE:R_D}
\subsection{Parameter configuration and candidate phases}
\begin{figure}[htbp!]
\centering
\includegraphics[width= 1\linewidth]{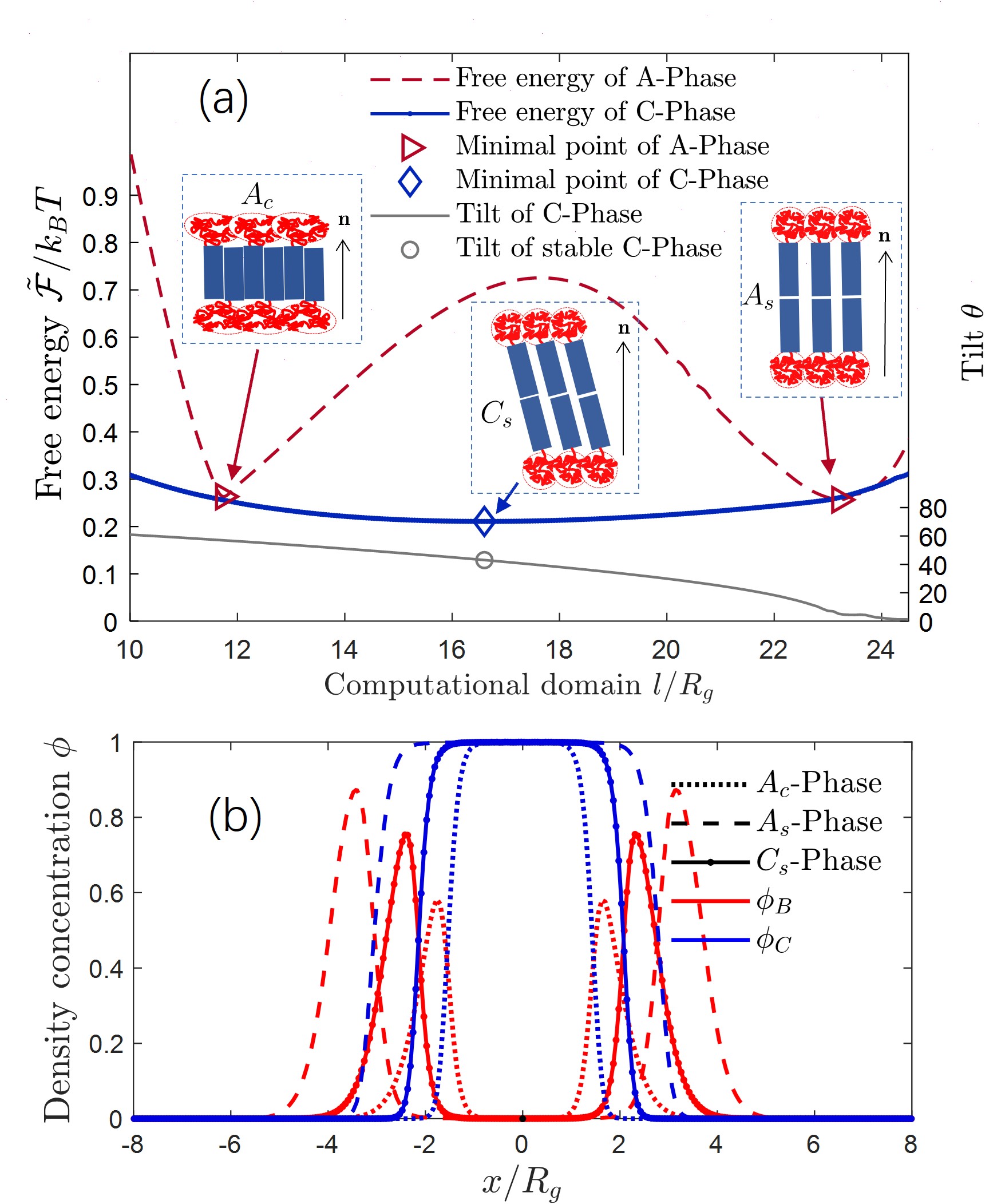}
\caption{
(a) The free energy $\tilde{\F}$ (left ordinate) of the liquid crystal A-phase bilayer and C-phase bilayer as functions of the computation domain $l$ in the canonical ensemble. Showns is the tilt $\theta$ (right ordinate) of the C-Rod (\emph{i.e.}, phospholipid tails) in the C-phase bilayer as a function of the computation domain $l$. Here, $f_B = 0.25$, $\beta_C = 4$, and $\varphi_{DE} = 0$. Tilt is obtained from the orientational density distribution of C-Rod $\S_C(x)$. Details are provided in the Appendix~\ref{App: tilt}.
(b) The density concentration $\phi$ of the steady state $A_c$-Phase, $A_s$-Phase, and $C_s$-Phase,
The red and blue curves show $\phi_B$ and $\phi_C$ respectively (\emph{i.e.}, phospholipid headgroup and phospholipid tail.)}
\label{fig:Aphase_Cphase}
\end{figure}

\textbf{Model parameters.}
In previous work \cite{wang2023mimicking}, we used a self-consistent field theory and provided model parameters for cholesterol without a headgroup. In this study, we improve the theory by modeling cholesterol as a diblock copolymer. The inclusion of a headgroup necessitated some significant deviations from a headless cholesterol.  In particular, we refined the model parameters by incorporating the volume and length ratios between the phospholipid and cholesterol headgroups. This approach offers two key advantages. First, considering the volume ratio between the cholesterol and phospholipid headgroups explicitly accounts for the much smaller headgroup of cholesterol than of a phospholipid. Second, it naturally leads to a C-phase. In this phase, the phospholipids are tilted from the normal in the absence of cholesterol. The C-phase corresponds to the $L_{\beta^{'}}$ phase experimentally observed in phospholipid bilayer membranes \cite{marrink2005simulation}, and we found that this phase has a lower free energy than the A-phase (see Fig.~\ref{fig:Aphase_Cphase}(a)).

\begin{figure}[htbp!]
\centering
\includegraphics[width= 1\linewidth]{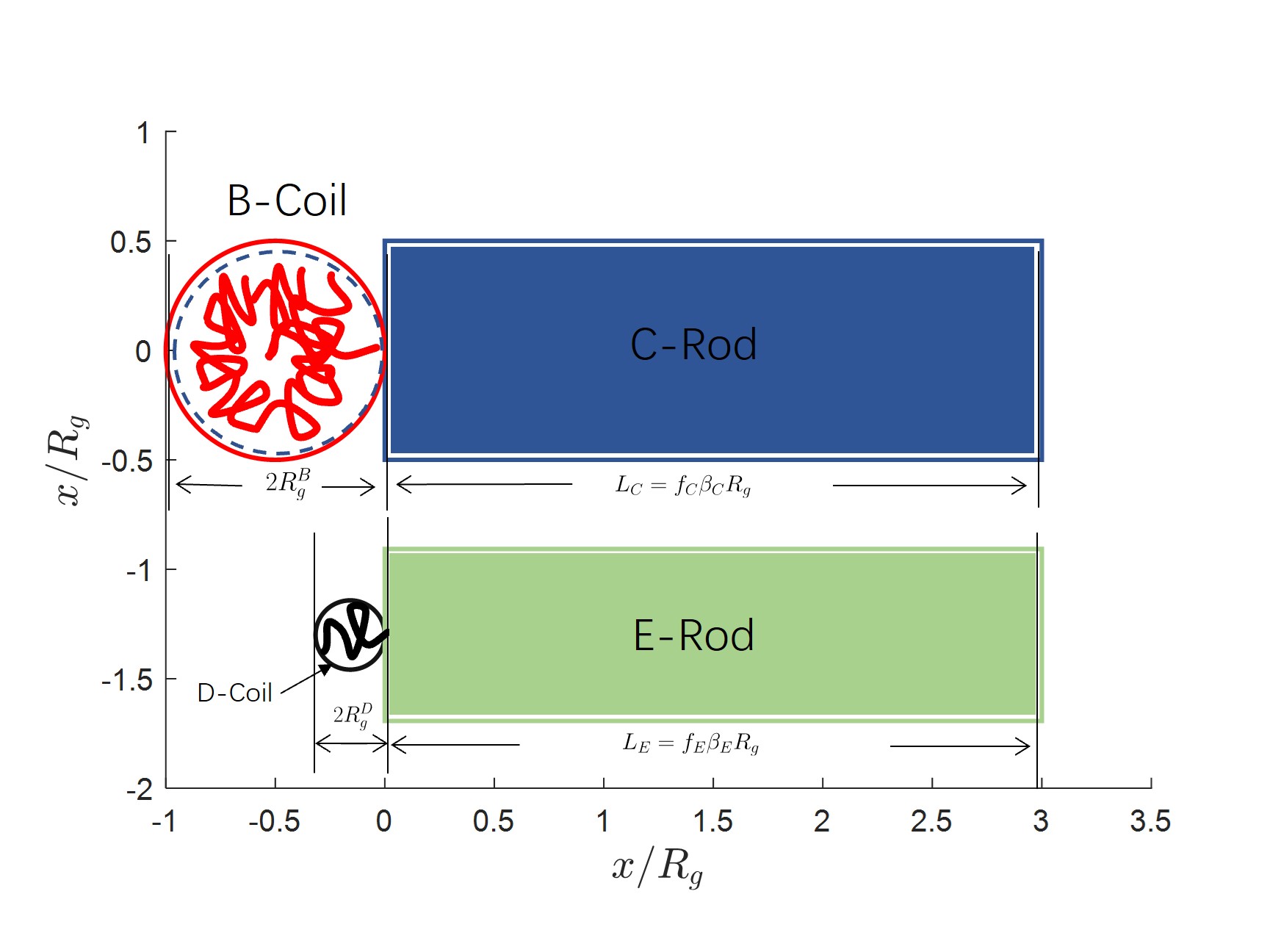}
\caption{Polymer size diagram with $f_B = 0.25$, $f_D = 0.025$, $f_C= f_E=0.75$, $\beta_C= \beta_E= 4$. A-Coil simulates water, BC-Rod-Coil simulates saturated phospholipids, and DE-Rod-Coil simulates cholesterol. The unit length is $R_g$. The red circle (B-Coil) and black circle (D-Coil) represent the headgroups of phospholipids and cholesterol, respectively. The blue rectangle (C-Rod) and green rectangle (E-Rod) represent the hydrophobic regions of a phospholipid and cholesterol, respectively.}
\label{fig:ABCDE_SIZE}
\end{figure}

The B-Coil segment can be approximated as a sphere with a radius of $R_g^B$. The diameter of this sphere is given by $L_B = 2R_g^B = 2 \sqrt{f_B} R_g$. Typically, the volume and length ratios between the headgroups and tails of phospholipids, $f_B:f_C$, range from 1:3 to 1:4 \cite{uhrikova2007component, miyoshi2014detailed}. In this work, we typically set $f_B : f_C = 1:3$, letting $f_B = 0.25$ and $f_C = 0.75$. The volume ratio between the phospholipid headgroup and cholesterol is approximately $1:10 \sim 1:20$; we let $f_B: f_D = 10:1$, leading to $f_D = 0.025$. Since the hydrophobic component of cholesterol is modeled as a single rod molecule, the length of the D-Rod corresponds to the cholesterol tail length, which we set equal to the length of the phospholipid tail. That is, $f_E = f_C$ and $L_E = L_C = f_C \beta_C$. The assumed sizes of volume fraction for the BC-Rod-Coil and DE-Rod-Coil components are illustrated in Fig.~\ref{fig:ABCDE_SIZE}.

After the scale parameters are determined, the Flory-Huggins and the Maier-Saupe interaction parameters must be set.  The ten Flory-Huggins parameters and single  Maier-Saupe parameter capture the fundamental characteristics of the bilayer membrane. The large number of parameters poses challenges for a multi-component polymer system. We based the values of the Flory–Huggins parameters on previous studies \cite{cai2017liquid,wang2023mimicking}.

The Flory-Huggins parameters were divided into two groups: attractive interactions, including $\chi_{AB} N$, $\chi_{AD} N$, and $\chi_{BD} N$, and repulsive interactions, including $\chi_{AC}N$, $\chi_{AE}N$, $\chi_{BC}N$, $\chi_{BE}N$, $\chi_{CD}N$, $\chi_{CE}N$, and $\chi_{DE}N$. Specifically, $\chi_{AB} N$ and $ \chi_{AD} N$ represent the interactions between water and the phospholipid and cholesterol headgroups, respectively. For simplicity, we let $\chi_{AB} N=\chi_{AD} N=\chi_{BD} N=0$. $\chi_{AC}N = \chi_{AE}N = 30$ to reflect the repulsion between water and phospholipid and cholesterol tails. Due to the small volume fraction of the cholesterol headgroup (D-Coil), repulsive interactions involving the D-Coil are set to slightly higher values, with $\chi_{CD}N = \chi_{DE}N = 40$.

To account for the attractive interaction between the phospholipid and cholesterol headgroup, $\chi_{BD}$ is set to $-30$. The distance-dependent interaction between the phospholipid and cholesterol tails has not been determined. We chose a moderate interaction parameter, $\chi_{CE} N = 0$. Since both phospholipid and cholesterol tails exhibit strong liquid crystalline order, the Maier-Saupe parameter is set to $\eta N = 30$. These default parameters, summarized in Table~\ref{tab:model parameter2}, were used, unless stated otherwise.

\renewcommand\arraystretch{1.5}
\newcolumntype{Y}{>{\centering\arraybackslash}X}
\begin{table*}[htbp!]
\caption{List of default parameters for the phospholipid cholesterol bilayer model.}
\label{tab:model parameter2}
\begin{center}
\begin{tabular}{  r  r  r  r  r  r  r}
\hline
&\multicolumn{2}{c}{\makecell{Number of polymer\\(unit $n_{BC}$)}}
&\multicolumn{1}{c}{}
&\multicolumn{3}{c}{\makecell {Volume fraction\\(unit $N$)}}
\\  \cline{2-3}\cline{5-7}
\makecell{Parameter \quad}
&\makecell{$\n_{A}  $}
&\makecell{$\n_{DE} $ }
 &
&\makecell{\quad$f_{B}$}
&\makecell{\quad$f_{D}$}
&\makecell{\quad$f_{C} = f_{E}$}
\\ \hline
\makecell{Value }  &  \makecell{2}   &\makecell{0 $\sim$ 1}    &  & \makecell{\quad0.25}& \makecell{\quad0.025} & \makecell{\quad0.75}

\\  \hline
&\multicolumn{3}{c}{\makecell{Asymmetric parameters}}
&\multicolumn{1}{c}{}
&\multicolumn{2}{c}{\makecell {Maier-Saupe }}
\\  \cline{2-4}\cline{6-7}
\makecell{Parameter \quad}
&\makecell{$\beta_{C}$}
&
&\makecell{$\beta_{E}$}
&
&\multicolumn{2}{c}{\makecell{$\eta N $ }}
\\ \hline
\makecell{Value}
&\makecell{4}
&
&\makecell{4}
&
&\multicolumn{2}{c}{\makecell{30}}
\\ \hline

&\multicolumn{6}{c}{\makecell {Flory-Huggins }}
\\  \cline{2-7}

\makecell{Parameter \quad}
&\multicolumn{3}{c}{ \makecell{$\chi_{AB}N = \chi_{AD}N = \chi_{BE} = 0$}}
&\multicolumn{3}{c}{\makecell{$\chi_{AC}N = \chi_{AE}N =\chi_{BC}N $}}
\\ \hline
\makecell{Value}
&\multicolumn{3}{c}{ \makecell{0}}
&\multicolumn{3}{c}{ \makecell{30}}
\\ \hline
&\multicolumn{6}{c}{\makecell {Flory-Huggins }}
\\  \cline{2-7}

\makecell{Parameter \quad}
&\multicolumn{2}{c}{ \makecell{$\chi_{BD}N$}}
&\multicolumn{1}{c}{\makecell{$\chi_{CD}N = \chi_{DE}N$ }}
&\multicolumn{3}{c}{\makecell{$\chi_{CE}N $}}
\\ \hline
\makecell{Value}
&\multicolumn{2}{c}{ \makecell{-30}}
&\multicolumn{1}{c}{ \makecell{40}}
&\multicolumn{3}{c}{ \makecell{0}}
\\ \hline
\end{tabular}
\end{center}
\begin{flushleft}
\footnotesize\textbf{Note:} Cholesterol concentration $\varphi_{DE} = \n_{DE}/(\n_{BC} + \n_{DE})$.
\end{flushleft}
\end{table*}

\begin{figure}[htbp!]
\centering\label{fig: Candidatephases}
\includegraphics[width= 1\linewidth]{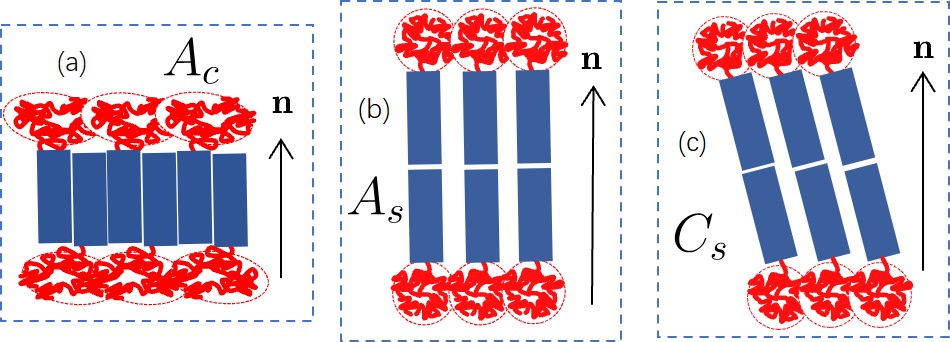}
\caption{The initial states used for simulating phospholipid-cholesterol bilayer membranes are a self-assembled Rod-Coil/Coil self-assembled system as described below, where $\mathbf{n}$ represents the normal direction of the bilayer membrane:
(a) In the $A_c$ phase bilayer membrane, the tails of the bilayer membrane are interdigitated, while the Rod molecules maintain their average orientation parallel to the normal direction of the bilayer membrane.
(b) In the $A_s$ phase bilayer membrane, Rod molecules (depicted in blue) are arranged in an end-to-end configuration, with their average orientation parallel to the normal direction of the bilayer membrane.
(c) In the $C_s$ phase bilayer membrane, Rod molecules are tilted from the normal direction of the bilayer membrane.}
\label{fig:reference_phase}
\end{figure}

\textbf{Candidate phases.}
We considered three different phases and determined which had the lowest free energy. For two of the phases, the lipids were normal to the membrane-solution interfaces (Fig.~\ref{fig: Candidatephases}). The lipids in the two monolayers could either interdigitate maximally ($A_c$) or not at all ($A_s$). Alternatively, the lipids could deviate from the normal by tilting (phase $C_s$). Scale parameters were selected from the volume ratio and length ratio of DPPC to cholesterol.

For pure phospholipid bilayers in a gel phase ($L_{\beta^{'}}$), the lipids are tilted away from the bilayer normal. This arrangement results in a more compact structure with reduced free energy. We represented the two saturated acyl chains of a phospholipid by a rod. Fig.~\ref{fig:Aphase_Cphase}(a) shows the free energy of pure phospholipid bilayer (\emph{i.e.} $\varphi_{DE} = 0$) in the $C_s$-phase and A-phase bilayers, with fixed parameters as functions of the computation domain $l$ (see Appendix~\ref{App: Computation domain}). In our model, calculations show that the gel phase has the lowest free energy. When simulating phospholipid-cholesterol bilayer membranes with a Rod-Coil/Rod-Coil/Coil polymer system, the $C_s$-phase provides an ideal reference state.

The free energy of the A-phase bilayer exhibits two local minima, $\tilde{\F} = 0.2633$ and $\tilde{\F} = 0.2566$ (indicated by triangles on the red dashed curve, Fig.~\ref{fig:Aphase_Cphase}(a)), corresponding to the metastable $A_c$- and $A_s$-phases, respectively. These minima osculate the free energy curve of the C-phase. It follows that the $A_c$- or $A_s$-phase is metastable, each requiring its own computation domain. The optimal computational domain is $l^{*}/R_g = 11.74$ for the $A_c$-phase and $l^{*}/R_g = 23.22$ for the $A_s$-phase. The optimal computational domain for the $A_s$-phase is approximately twice that of the $A_c$-phase. This indicates that the interdigitation of the C-Rods decreases with increased computation domain. Fig.~\ref{fig:Aphase_Cphase}(a) provides an energy landscape of A-phase and C-phase bilayers and visually depicts their structures.

It is notable that the gel phase always has the lowest free energy.  The free energy of the $C_s$-phase bilayer membrane reaches its minimum at $l^{*}/R_g = 16.6$, where the stable $C_s$-phase has a free energy of $\tilde{\F} = 0.2108$. This is the equilibrium state. As the computation domain increases, the tilt of the C-Rod molecules in the C-phase bilayer gradually decreases (Fig.~\ref{fig:Aphase_Cphase}(a), gray curve). At equilibrium, the average tilt of the $C_s$-phase is approximately $43^{\circ}$ (indicated by circle on the gray curve, Fig.~\ref{fig:Aphase_Cphase}(a)). This value is larger than the experimentally measured $30^{^\circ}$ \cite{katsaras1992fatty,sun1994order}. Although there are numerical differences from experimental determinations, our simulation captures the essential qualitative features of phospholipid tilt and identifies the $C_s$-phase as the most stable state.

Fig.~\ref{fig:Aphase_Cphase}(b) shows the density distributions of the metastable structures ($A_c$- and $A_s$-phases) and the stable structure ($C_s$-phase) of cholesterol-free bilayers, represented by dotted, dashed, and solid curves, respectively. These density distributions reveal differences between the three phases. Among them, the $A_c$-phase exhibits the smallest membrane thickness, with a hydrophobic core thickness of approximately $2.925R_g$. This small thickness is clearly due to the interdigitation of the phospholipid tails. The $A_s$-phase displays the greatest membrane thickness, approximately $5.775R_g$. This doubling of thickness is as expected for non-interdigitated \emph{vs} and. interdigitated acyl chains. T.  The tilted arrangement of phospholipids in the $C_s$-phase results in a bilayer thickness of approximately $4.16R_g$. Analysis of the thickness of the three phases provides an alternate way to estimate the tilt of the equilibrium state $C_s$-phase. Lipid tilt in the $C_s$-phase bilayer based on the density distrbution is approximately $180^{\circ} \arccos((4.16R_g)/(5.775R_g))/\pi \approx 44^{\circ}$. This value is very close to the tilt of $43^{\circ}$, obtained from Fig.~\ref{fig:Aphase_Cphase}(a), gray circle, for the computation domain of $l = 16.6R_g$.

\subsection{The effect of cholesterol concentration}
In this section, we investigate the ability of the $C_s$-phase to capture the effects of cholesterol concentration on the bilayer membrane. Explicitly, we calculated cholesterol chemical potentials to thermodynamically evaluate the model,  and lipid tilt and membrane thickness to test the model's ability to capture structural aspects. Additionally, we examined whether the model predicted the classical condensation effect of cholesterol.

\subsubsection{Chemical potential of cholesterol}

\begin{figure}[htbp!]
\centering
\includegraphics[width= 1\linewidth]{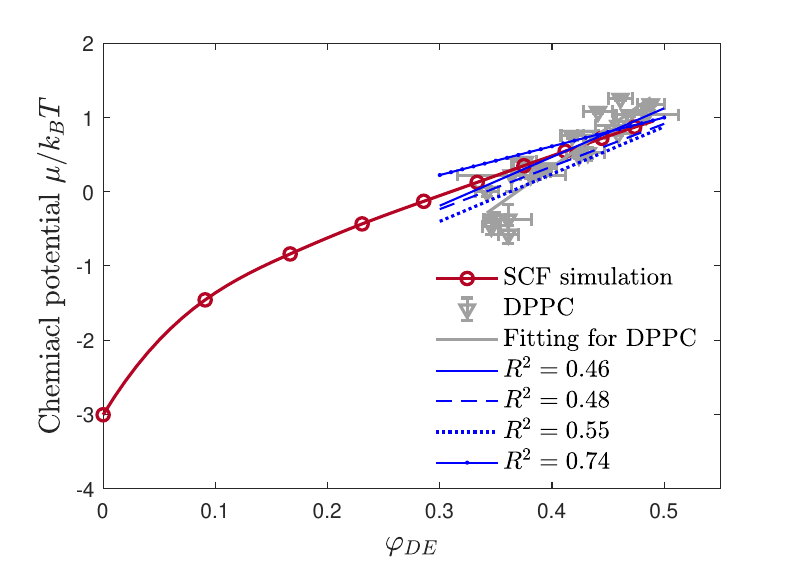}
\caption{Comparison of chemical potentials. The parameters for the chemical potentials obtained from self-consistent field simulations are listed in Table~\ref{tab:model parameter2}. The grey triangles represent the experimental measurements of the cholesterol chemical potential in the DPPC bilayer. The curve with $R^2 = 0.46$ corresponds to fitting $\mu$ to lattice models of pairwise interactions between components. The curve with $R^2 = 0.48$ represents fitting $\mu_c$ to condensed complexes where cholesterol forms complexes with DPPC. The curve with $R^2 = 0.55$ extends this model by incorporating an additional repulsive interaction between the complexes and DOPC. Finally, the curve with $R^2 = 0.74$ represents fitting $\mu_c$ to mean-field models of pairwise interactions between components. The experimental and fitting data are sourced from Ref.~\cite{shaw2023chemical}. The chemical potentials were adjusted so that their values at a cholesterol concentration of $30\%$ were set to zero.}
\label{fig:5_SCF_DPPC}
\end{figure}

An approach developed to quantitatively measure the chemical potential and activity of cholesterol within membranes of erythrocytes and nucleated cells \cite{ayuyan2018chemical} has been applied to liposomes composed of various mixtures of phospholipids and cholesterol \cite{shaw2023chemical}. Fig.~\ref{fig:5_SCF_DPPC} presents the cholesterol chemical potential obtained from self-consistent field simulations (see Appendix~\ref{App: CH_DE} for the equations used to obtain chemical potential). This chemical potential exhibits a continuous, monotonic increase with cholesterol concentration, and is quantitatively in good agreement with experimental measurements of the cholesterol chemical potential in DPPC /cholesterol bilayers. When the cholesterol concentration is below $0.3$, the chemical potential varies nonlinearly with cholesterol concentration. At higher cholesterol concentrations, within the range of $0.37$ to $0.5$, the chemical potential varies linearly with cholesterol concentration. The gray solid curve in Fig.~\ref{fig:5_SCF_DPPC} is the linear fit to the experimental data for the DPPC bilayer. The experimental chemical potentials between concentrations of $0.32$ and $0.37$ appear to be a multivalued function of cholesterol concentration. We assume this is a consequence of experimental noise and is not real.

Fig.~\ref{fig:5_SCF_DPPC} also displays chemical potential values, $\mu$, derived from a variety of models used to account for the liposome experimental data \cite{shaw2023chemical}; the Pearson $R^2$ values of $0.46$, $0.48$, $0.55$, and $0.74$. are shown in the figure. The fitting curves of these models closely align with our simulation results, all exhibiting a consistent linear trend at high cholesterol concentrations. The self-consistent field model appears to offer a robust framework for deriving thermodynamic properties of lipid bilayers.

\begin{figure}[htbp!]
\centering
\includegraphics[width= 0.8\linewidth]{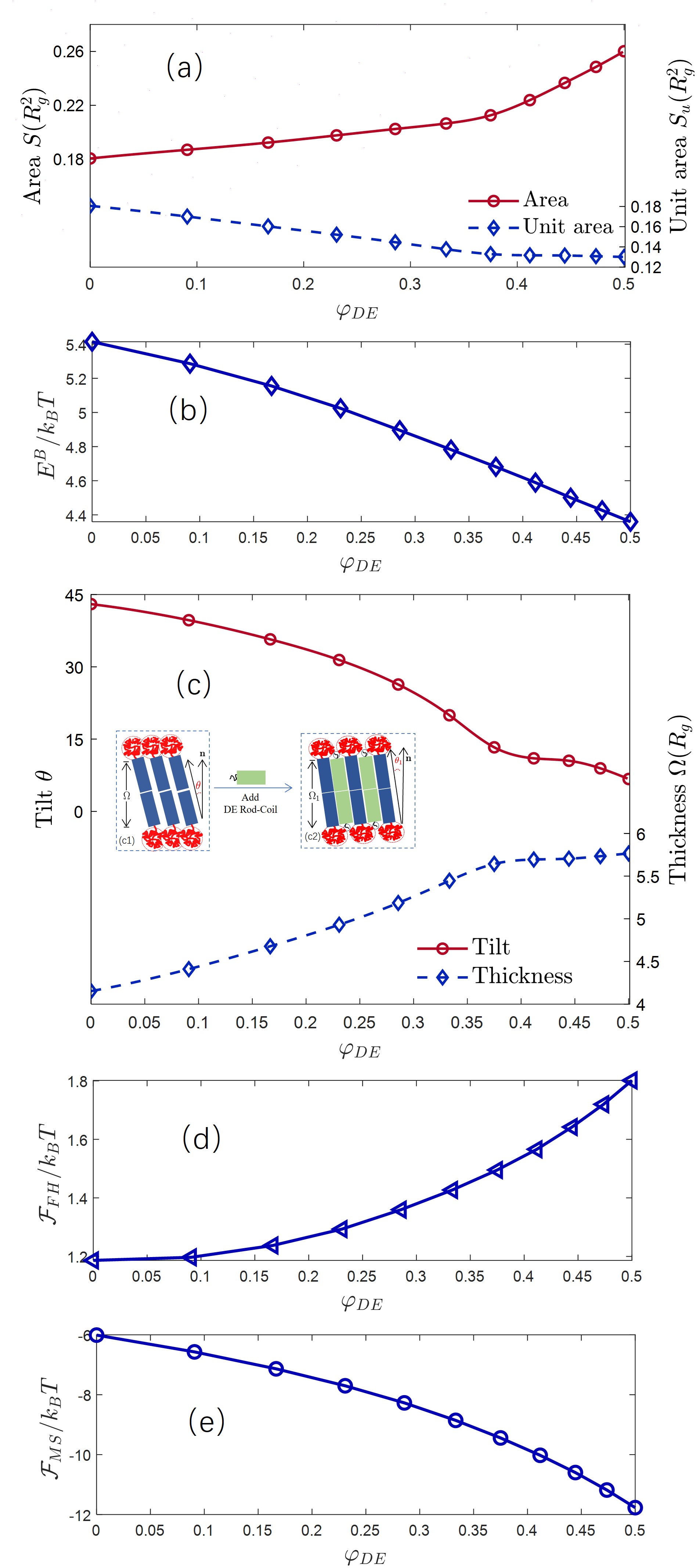}
\caption{
(a) The bilayer area $S$ (left ordinate, solid curve) and the unit area $S_u$
 (right ordinate, dashed curve)  as a function of cholesterol concentration $\varphi_{DE}$. Note: $S_u = S/(\n_{BC} + \n_{DE})$.
(b) The stretching entropy loss $E^{B}$ of the B-Coil as a function of the cholesterol concentration $\varphi_{DE}$. The equations for the stretching entropy loss $E^{B}$ are presented in Appendix~\ref{App:Energy classification}.
(c) The tilt $\theta$ (left ordinate, solid curve) of C-Rod and the bilayer thickness $\Omega$ (right ordinate, dashed curve)  as a function of the cholesterol concentration $\varphi_{DE}$.
(Inset c1) Schematic diagram of the $C_s$-phase bilayer membrane without DE-Rod-Coils ($\varphi_{DE}= 0$), where $\theta$ is the average tilt of C-Rod tails and $T$ is the thickness of the bilayer membrane.
(Inset c2) Schematic diagram of the $C_s$-phase bilayer membrane after adding DE-Rod-Coil, where $\theta_1$ is the average tilt of C-Rod tails and $\Omega_1$ is the thickness of the bilayer membrane. Here, $\theta > \theta_1$, and $\Omega < \Omega_1$.
(d) The Flory-Huggins interaction potential of BC copolymers and DE copolymers as functions of the cholesterol concentration $\varphi_{DE}$.
(e) The Maier-Saupe interaction potential of C-Rod and E-Rod as functions of the cholesterol concentration
$\varphi_{DE}$.  The equations for obtaining the stretching entropy loss, Flory-Huggins potentials, and  Maier-Saupe potentials are provided in Appendix~\ref{App:Energy classification}.
}
\label{fig:Thckness_Angel}
\end{figure}

\subsubsection{The condensation effect}

Among the significant effects of greater cholesterol content on a phospholipid bilayer membrane are increased lipid packing with a consequent decrease in fluidity, increased elastic moduli, and decreased membrane leakage \cite{rog2001cholesterol,boughter2016influence}. More classically, known for $100$ years, is the condensation effect: upon cholesterol addition, the area of a bilayer or monolayer composed of saturated phosphatidylcholines(PCs) increases less than the area of the added cholesterol alone \cite{leathes1925croonian}. In other words, the area of a saturated PC/cholesterol membrane is less than the sum of the areas that would be contributed by the PC and cholesterol alone. We previously used homopolymers to model the effect of a headless cholesterol on bilayer membrane, but the model failed to predict the condensation effect.

We now report that self-consistent field simulations for block copolymers do predict the condensation effect (Fig.~\ref{fig:Thckness_Angel}(a)). The total bilayer area $S$ increases with increasing $\varphi_{DE}$, but the average area per polymer (referred to as the unit polymer area) $\tilde S$ of the bilayer decreases with increased $\varphi_{DE}$. The decrease in area per polymer is a consequence of changes in the distribution of headgroups. Specifically, as cholesterol increases in the bilayer, the density distribution of phospholipid headgroups also increases along the tangential direction of the bilayer membrane interface. This increased density more effectively shields cholesterol, with its small headgroup, from water. We presume that this redistribution requires the phospholipid headgroups to stretch with increased cholesterol concentration. The increased density intrinsic to the condensation effect results in an entropic penalty, referred to as an entropy loss, because of reduced degrees of freedom. This entropy loss (Fig.\ref{fig:Thckness_Angel}(b)) is thermodynamically unfavorable (See Appendix~\ref{App:Energy classification} for the equations used to obtain entropy loss). The favorable enthalpic gain in energy resulting from greater attractive interaction energy at higher lipid density more than compensates for the entropic loss. The predictions of our model are consistent with experimental measurements \cite{hung2007condensing}. The finding that lipid headgroups shield cholesterol from water, and that this accounts for the condensation effect, is similar to the explanation provided by the umbrella model \cite{huang1999microscopic}. However, our approach allows explicit calculation and provides quantitative results. For example, the model predicts that the percentage of condensation (\emph{i.e.}, the reduction in unit area) remains linear up to a cholesterol concentration of $37\%$; this is higher than the  $30\%$ concentration observed in molecular dynamics simulations \cite{pan2012interactions}.

\subsubsection{Bilayer thickness and lipid tilt}

Fig.~\ref{fig:Thckness_Angel}(c) illustrates phospholipid tilt and bilayer thickness as a function of cholesterol concentration. As the cholesterol concentration increases to $\varphi_{DE} = 0.375$, the phospholipid tilt decreases progressively from $\theta \approx 43^{\circ}$ in the absence of cholesterol to $\theta \approx 13^{\circ}$.

As more cholesterol is added to the bilayer, the cross-sectional area of the phospholipid headgroups increases: they (\emph{i.e.}, B-Coils) stretch to cover the hydrophobic core of cholesterol from the water (\emph{i.e.}, A-Coil), as described above. The cross-sectional areas of the headgroups and the tails collectively determine the degree of tilt of the tails. The less the phospholipids tilt, the less their headgroups have to stretch to sequester cholesterol from water. Thus, as the cholesterol concentration increases, the tilt of the phospholipid tails decreases (Fig.~\ref{fig:Thckness_Angel}(c)). As tilt decreases, the tails become more ordered, as seen from the decrease in the Maier-Saupe potential (Fig.~\ref{fig:Thckness_Angel}(e)). Additionally, in our simulations, the effect of the Maier-Saupe potential plays a dominant role (\emph{i.e.}, the  orientation interactions between the polymers are a dominant factor), as evidenced by the difference in energies for $\varphi_{DE} = 0$ and $\varphi_{DE} = 0.5$: it is only $~0.6$ for $\F_{FH}$ (Fig.~\ref{fig:Thckness_Angel}(d)), but an order of magnitude greater, $~6$, for $\F_{MS}$ (Fig.~\ref{fig:Thckness_Angel}(e). In short, a decrease in phospholipid tilt with increased cholesterol reduces the total free energy of the system. (See Appendix~\ref{App:Energy classification} for the equations used to obtain $\F_{FH}$ and $\F_{MS}$.)

At higher cholesterol concentrations ($\varphi_{DE} = 50\%$), the phospholipid tilt stabilizes at a minimal value of $\theta < 10^{\circ}$, a small tilt. At this concentration, the bilayer structure approaches that of the $A_s$-phase. The reduction in phospholipid tilt directly causes the bilayer thickness to increase (Fig.~\ref{fig:Thckness_Angel}(c)). Specifically, as the phospholipid tilt decreases from $\theta$ to $\theta_1$, the bilayer thickness increases from $\Omega$ to $\Omega_1$, as illustrated in the inset of Fig.~\ref{fig:Thckness_Angel}(c). Bilayer thickness grows continuously and almost linearly from $\Omega \approx 4.16 R_g$ to $\Omega \approx 5.64 R_g$ as the cholesterol concentration rises from $\varphi_{DE} = 0$ to $\varphi_{DE} = 0.375$. Over the cholesterol concentration range of $\varphi_{DE} = 0.375 \sim 0.5$, phospholipid tilt hardly varies, and, accordingly, thickness remains relatively constant.

In summary, as cholesterol concentration increases from $0\%$ to $50\%$, the unit lipid area, bilayer thickness, phospholipid tilt, and chemical potential of cholesterol exhibited monotonic changes that were in excellent qualitative accord with experimental data\cite{hung2007condensing,pan2008cholesterol,shaw2023chemical}.

\subsection{The effects of volume fraction and hydrophobic length}

\subsubsection{Volume fraction}
\begin{figure}[htbp!]
\centering
\includegraphics[width= 1\linewidth]{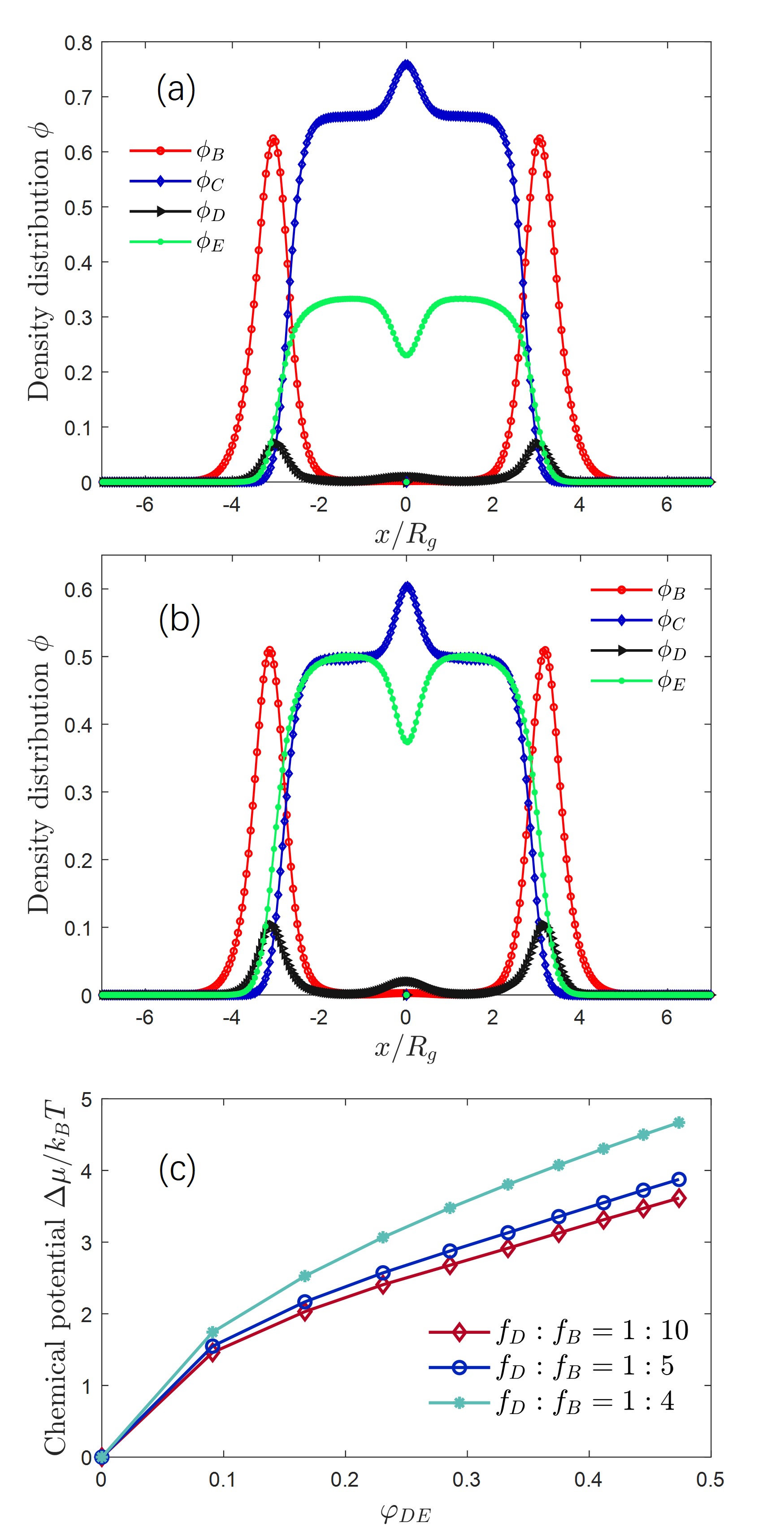}
\caption{The density distributions of a bilayer for two cholesterol concentrations. (a) $\varphi_{DE} = 0.3333$, and (b) $\varphi_{DE} = 0.5$. (c) The cholesterol chemical potential $\mu_{DE}$ of the bilayer under different volume
fractions ($f_D:f_B = 1:10$, $f_D:f_B = 1:5$, and $f_D:f_B = 1:4$) as functions of cholesterol concentration $\varphi_{DE}$, with $f_B$  fixed at $0.25$.}
\label{fig:density_profile}
\end{figure}

Fig.~\ref{fig:density_profile}(a) and Fig.~\ref{fig:density_profile}(b) show the normal density distributions of bilayer membranes for two cholesterol concentrations: $\varphi_{DE} = 0.3333$ (Fig.\ref{fig:density_profile}(a)) and $\varphi_{DE} = 0.5$ (Fig.\ref{fig:density_profile}(b)). Because the density of cholesterol must increase with an increase in its concentration, the density of the phospholipid tails must decrease, as governed by the incompressibility constraint, $\sum\limits_{i}\phi_i = 1$. At $\varphi_{DE} = 0.3333$, the integral ratio $\int_{-{l}/{2}}^{{l}/{2}}\phi_C(x)dx : \int_{-{l}/{2}}^{{l}/{2}}\phi_E(x)dx \approx 2:1$. This calculated ratio is precisely the phospholipid-to-cholesterol molar ratio of $\n_{BC}:\n_{DE} = 2:1$, illustrating the reliability of our calculations. Also, the relatively small volume occupied by the cholesterol headgroup results in a lower value of $\phi_D$ at small $f_D$. Because the phospholipid-to-cholesterol molar ratio is $\n_{BC}:\n_{DE} = 1:1$ for $\varphi_{DE} = 0.5$, here the density distribution of the phospholipid tails and cholesterol hydrophobic core becomes similar Fig.~\ref{fig:density_profile}(b).

Fig.~\ref{fig:density_profile}(c) shows the chemical potential of cholesterol as a function of cholesterol concentration under several volume ratio conditions ($f_D:f_B = 1:10$, $f_D:f_B = 1:5$, and $f_D:f_B = 1:4$). An increasing volume ratio of cholesterol headgroup to phospholipid headgroup results in a steeper change in chemical potentials with cholesterol concentration. This occurs because a higher volume fraction of a cholesterol headgroup corresponds to a greater degree of polymerization of the headgroup. The ability of headgroups to have greater stretching/contracting immediately augments the repulsive and attractive interactions experienced by the cholesterol headgroup. However, repulsion between the D-Coil and other polymers contributes more significantly to the membrane’s free energy.

\subsubsection{Hydrophobic length}
The energy from hydrophobic interactions is directly dependent on membrane thickness \cite{aranda2001electromechanical,itel2014molecular}. So far, we have maintained the same hydrophobic length for phospholipids and cholesterol: $L_C = f_C\beta_C = 0.75 * 4  = 3R_g$ and $L_C = f_C\beta_C = 0.800625 * 4.1  = 3.2826R_g$. We altered the strength of hydrophobic interactions by changing the length of the phospholipid tails, but maintaining the length of cholesterol at $L_E = f_E\beta_E = 3R_g$.

\begin{figure}[htbp!]
\centering
\includegraphics[width= 1\linewidth]{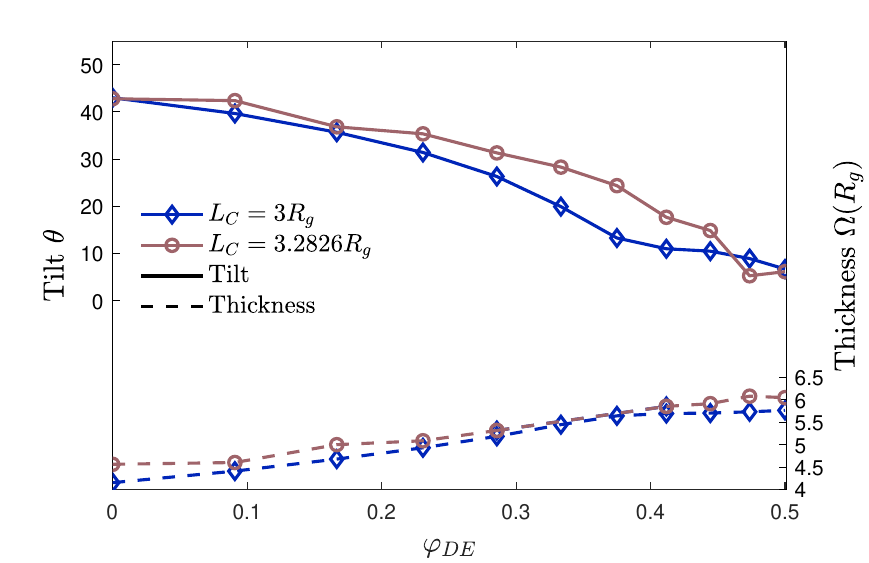}
\caption{The tilt $\theta$ (left ordinate, solid curve) of the C-Rod and the bilayer thickness $\Omega$ (right ordinate, dashed curve)  as a function of the cholesterol concentration $\varphi_{DE}$ for two phospholipid acyl chain lengths $L_C = 3R_g$ ($f_C = 0.75, \beta_C = 4R_g$), or $3.2826R_g$ ($f_C = 0.800625, \beta_C = 4.1R_g$).}
\label{fig:varlebgth}
\end{figure}

Fig.~\ref{fig:varlebgth} shows the impact of varying hydrophobic lengths of the diblock copolymers. Initially, we examine the effect on lipid tilt. The solid curves in Fig.~\ref{fig:varlebgth} present phospholipid tilt, for varying tail lengths of the phospholipid ($L_C = 3R_g$, $L_C = 3.2826R_g$) as a function of cholesterol concentration.

In bilayers composed exclusively of phospholipids, a moderate increase in the phospholipid tail length does not cause a significant change in the tilt: for $L_C = 3R_g$, and $L_C = 3.2826R_g$, the tilt is the same, $\theta = 42^{\circ}$. As cholesterol concentration increases, the tilt for $L_C = 3.2826R_g$ decreases substantially, to below $10^{\circ}$ at $45\%$ cholesterol. At this concentration, the bilayer structure resembles the A-phase, and tilt remains roughly constant even though cholesterol content increases. At the same cholesterol concentration, longer phospholipid tails result in greater tilt. For examble, at  $\varphi_{DE} = 0.3333$, when $L_C = 3.2828R_g$, the tilt is $\theta \approx 28^{\circ}$, and when $L_C = 3R_g$, the tilt is $\theta \approx 19^{\circ}$. In the presence of cholesterol, longer phospholipid tails ensure tighter alignment within the bilayer by adopting a larger tilt angle, promoting structural stability of the bilayer.

A decrease in tilt results in an increase in bilayer thickness (Fig.~\ref{fig:varlebgth}, right ordinate). For a length hydrophobic core, paramaterized by $L_C = 3R_g$, membrane thickness increases with cholesterol content, but does not further increase beyond $\varphi_{DE} \approx 0.375$. The same phenomenon holds for $L_C = 3.2826R_g$: greater thickness as cholesterol is enriched, with thickness saturation occurring at $\varphi_{DE} \approx 0.473$. The ranges of thickness increase with cholesterol content precisely match the ranges of tilt decreases. These findings show that once there is enough cholesterol to cause the structural transition from the C-phase to the A-phase, even more cholesterol affects neither tilt nor thickness.

\subsection{The effect of headgroup interactions}
The headgroup interactions $\chi_{AB} N$, $\chi_{AD} N$, and $\chi_{BD} N$,  correspond to the interactions between the solvent and the phospholipid headgroup, the solvent and the cholesterol headgroup, and the interaction between the phospholipid and cholesterol headgroups, respectively. So far, we set $\chi_{AB} N$ and $\chi_{AD} N$ to zero, minimizing their influence. We now relax these conditions in order to assess the impact of the headgroup interaction.

\subsubsection{Lipid tilt and chemical potential of cholesterol}
% \begin{figure}[htbp!]
% \centering
% \includegraphics[width= 1\linewidth]{figures/varinteraction_new_Angle.jpg}
% \caption{(a)The Tilt $\theta$ of lipids as functions of interactions $\chi_{AB}N$ with cholesterol concentrations  $\varphi_{DE} = 0,0.2,0.4$.
% (b)The Tilt $\theta$ of lipids as functions of interactions $\chi_{AD}N$ with cholesterol concentrations  $\varphi_{DE} = 0.2, 0.4$.
% (c)The Tilt $\theta$ of lipids as functions of interactions $\chi_{BD}N$ with cholesterol concentrations  $\varphi_{DE} = 0.2, 0.4$.}
% \label{fig:var_interaction_Angle}
% \end{figure}
% Phospholipid tilt, $\theta$, is shown as a function of headgroup interactions for three cholesterol concentrations: $\varphi_{DE} = 0$, $0.2$ or $0.4$ (Fig.\ref{fig:var_interaction_Angle}).

\begin{figure}[htbp!]
\centering
\includegraphics[width= 1\linewidth]{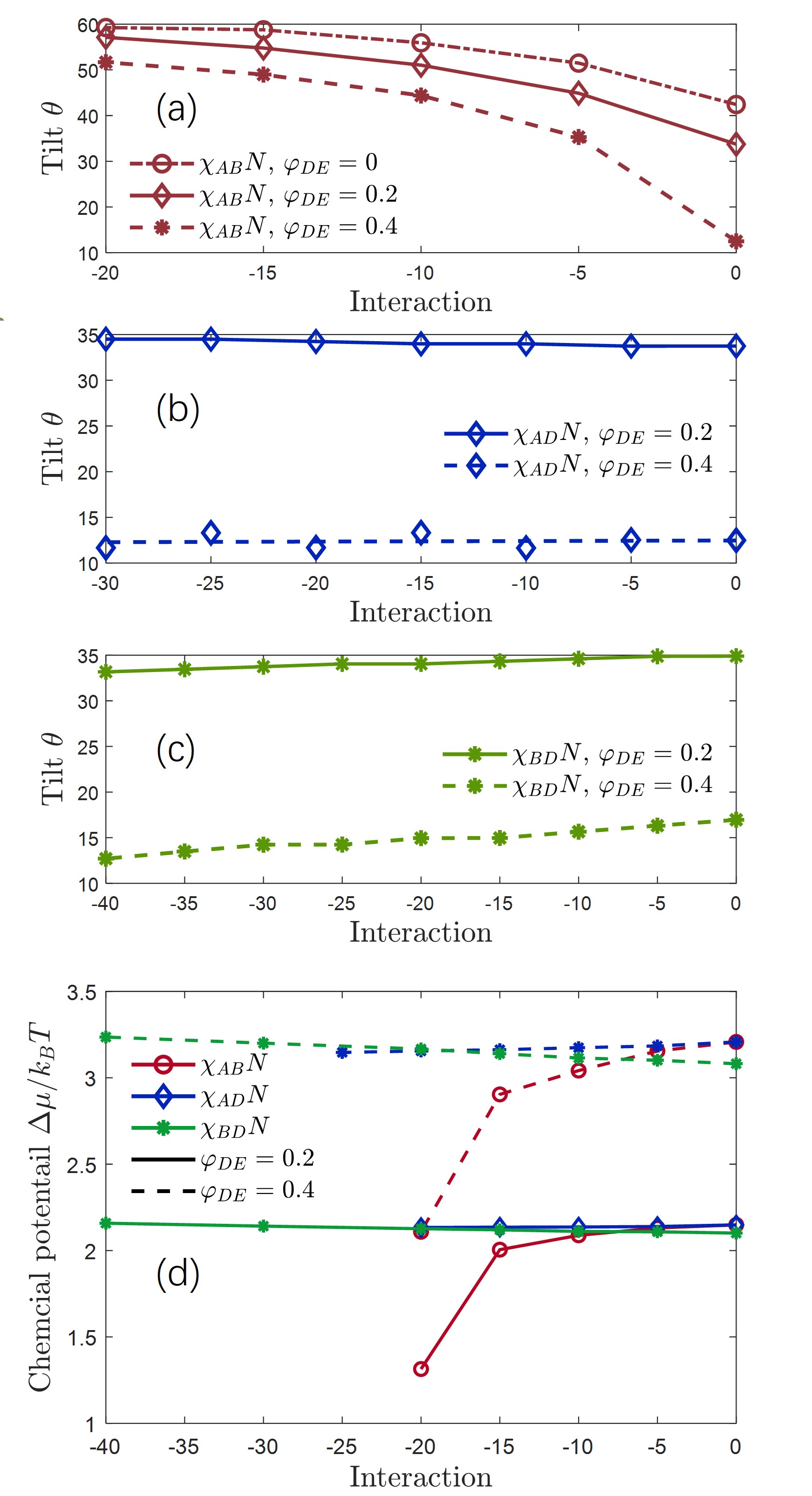}
\caption{
(a) The tilt $\theta$ of lipids as functions of interactions $\chi_{AB}N$ with cholesterol concentrations  $\varphi_{DE} = 0,0.2,0.4$.
(b) The tilt $\theta$ of lipids as functions of interactions $\chi_{AD}N$ with cholesterol concentrations  $\varphi_{DE} = 0.2, 0.4$.
(c) The tilt $\theta$ of lipids as functions of interactions $\chi_{BD}N$ with cholesterol concentrations  $\varphi_{DE} = 0.2, 0.4$.
(d) The chemical potential $\mu$ of cholesterol as functions of different headgroup interactions ($\chi_{AB}N, \chi_{AC}N, \chi_{AD}N$) with cholesterol concentrations of $\varphi_{DE} = 0.2$ (solid curve) and $\varphi_{DE} = 0.4$ (dashed curve).}
\label{fig:var_interaction}
\end{figure}

Phospholipid tilt significantly increases as the attractive interactions between the phospholipid headgroups and water increase, from $\chi_{AB}N = 0$ to $-20$ (Fig.~\ref{fig:var_interaction}(a)). The effect is nonlinear and holds for all cholesterol concentrations. Increasing cholesterol concentration blunts the effect. Increased attraction between headgroup and water causes the area of contact between the water and headgroups to increase. This requires that the unit polymer area, $S_u$, increase. Volume incompressibility requires that the thickness of the bilayer decrease, achieved through tilting of lipids away from the normal. Also, as energetically expected, increasing the attractive interaction $\chi_{AB}N $ from $0$ to $-20$ results in a marked decrease in the chemical potential of cholesterol, regardless of whether the cholesterol concentration is $\varphi_{DE} = 0.2$ or $\varphi_{DE} = 0.4$ (Fig.~\ref{fig:var_interaction}(d), red curve).

On the other hand, interactions between the cholesterol headgroup and either the solvent or the phospholipid head group have a negligible impact on phospholipid tilt. At a cholesterol concentration of $\varphi_{DE} = 0.2$, despite a significant strengthening of the attractive interactions between the phospholipid and cholesterol head groups ($\chi_{BD}N = 0 \sim -40$, Fig.~\ref{fig:var_interaction}(c)) or between the cholesterol head group and the solvent ($\chi_{AD}N = 0 \sim -30$, Fig.~\ref{fig:var_interaction}(b)), phospholipid tilt only slightly decreases.

% \begin{figure}[htbp!]
% \centering
% \includegraphics[width= 1\linewidth]{figures/varinteraction_new_CH.jpg}
% \caption{(The chemical potential $\mu$ of cholesterol as functions of interactions $\chi_{AB}N$(a), $\chi_{AC}N$(b), $\chi_{AD}N$(c) with cholesterol concentrations of $\varphi_{DE} = 0.2$(real line) and $\varphi_{DE} = 0.4$(dashed line). Choose the case where each zero cholesterol is the reference chemical potential.}
% \label{fig:var_interaction_CH}
% \end{figure}

At the higher cholesterol concentration of $\varphi_{DE} = 0.4$, the effects of $\chi_{BD}$ and $\chi_{AD}N$ on tilt are very slightly more consequential. The effect of $\chi_{BD}N$ on the phospholipid tilt is stronger than that of $\chi_{AD}N$. But again, the consequences are quite modest. This is closely associated with the volumes of water and cholesterol. In general, the outcomes of varying headgroup interactions can be physically understood within the context of the polymer model by considering the volume of a headgroup, and that increased attractive interactions cause the polymer headgroup to stretch; repulsion causes it to contract. In the simulation, we defined a ratio consistent with experimental measurement: a cholesterol headgroup $f_D=0.025$ and a phospholipid headgroup $f_B=0.25$ capture the relatively small volume of the cholesterol headgroup, and this leads to a correspondingly low density distribution (see Fig.~\ref{fig:var_interaction_density}(d)(e)). As a result, even when the attractive interactions related to the cholesterol headgroup are increased (\emph{i.e.}, variation $\chi_{BD}N, \chi_{AD}N$), the small headgroup cannot stretch enough to significantly increase its contact area with water or phospholipid headgroups. The negligible effect on the tilt of the C-Rod within the contact area immediately follows. The chemical potential of cholesterol also depends strongly on $\chi_{AB}N$, but only weakly on $\chi_{BD}N$ or $\chi_{AD}N$, for both cholesterol concentrations of $\varphi_{DE} = 0.2$ and $\varphi_{DE} = 0.4$ (Fig.~\ref{fig:var_interaction}(d)).

\subsubsection{Density concentration}
\begin{figure}[htbp!]
\centering
\includegraphics[width= 1\linewidth]{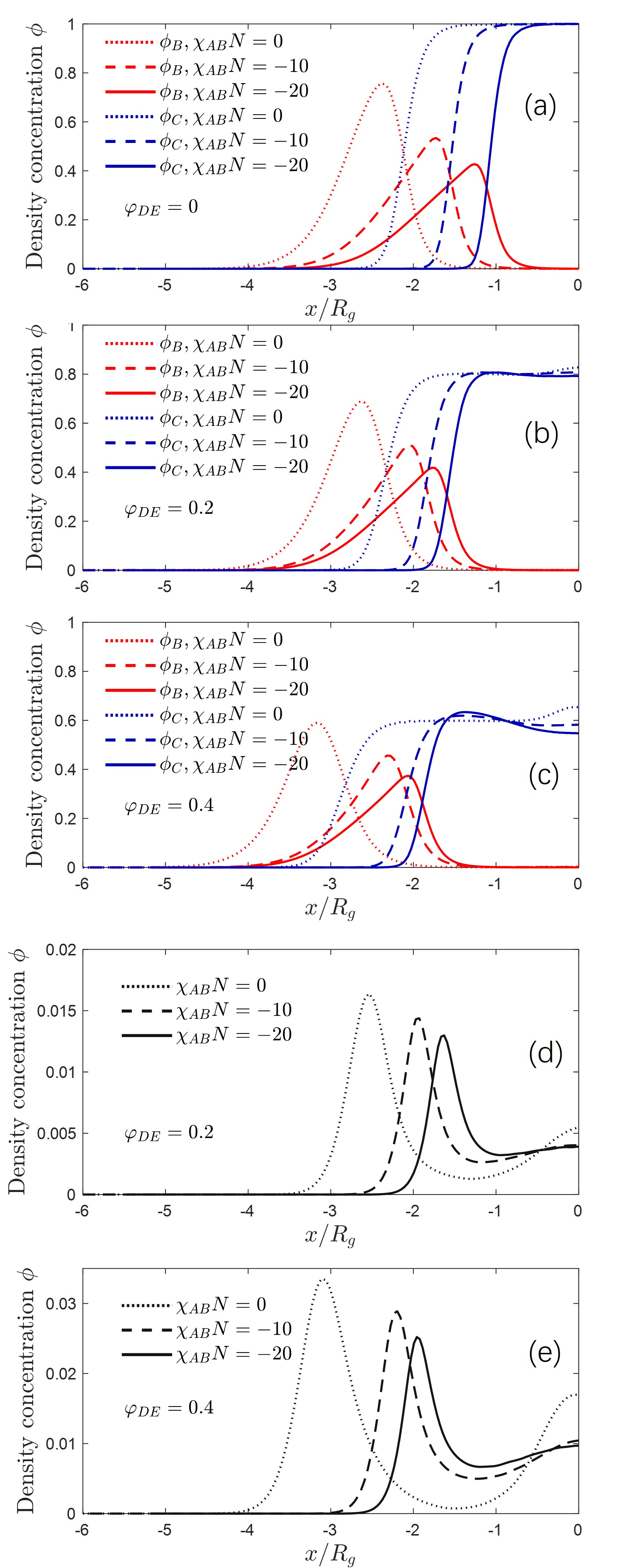}
\caption{The density concentration $\phi$ of the phospholipid for different values of $\chi_{AB} = 0$ (dotted curve), $-10$ (dashed curve), $-20$ (solid curve) at cholesterol concentrations of $\varphi_{DE} = 0$ (a), $\varphi_{DE} = 0.2$ (b), and $\varphi_{DE} = 0.4$ (c).
The density concentration $\phi$ of the cholesterol headgroup for different values of $\chi_{AB} = 0$ (dotted curve), $-10$ (dashed curve), $-20$ (solid curve) at cholesterol concentrations of  $\varphi_{DE} = 0.2$ (d) and $\varphi_{DE} = 0.4$ (e). (A negative Flory-Huggins parameter represents attraction) }
\label{fig:var_interaction_density}
\end{figure}

The underlying reason changes in $\chi_{AB}N$ strongly affect both phospholipid tilt and the cholesterol chemical potential can also be understood by appreciating that this Flory-Huggins interaction markedly affects the density distributions of headgroups and hydrophobic cores along the normal of the bilayer (Fig.~\ref{fig:var_interaction_density}).  Fig.~\ref{fig:var_interaction_density} (a), (b) and (c) illustrate, for different values of $\chi_{AB}N = 0$, $-10$, $-20$, the density distributions of phospholipids at cholesterol concentrations of $\varphi_{DE} = 0$, $\varphi_{DE} = 0.2$ and $\varphi_{DE} = 0.4$, respectively. The red curve shows the density distribution of the phospholipid headgroup. As the attraction between the phospholipid headgroup and the solvent increases, contact between the headgroup and water becomes more favorable, and so this area of contact increases. Because the bilayer is volumetrically incompressible, a greater area of contact necessitates that the thickness of the membrane decreases.  In other words, the distance decreases between the interfacial phospholipid headgroups and the center of the bilayer membrane. The phospholipid tails thus concentrate over a narrower range. At $\chi_{AB}N = 0$, the phospholipid tails are distributed in the range $x/R_g \in [-3.7,0]$, while at $\chi_{AB}N = -20$, this range is reduced to $x/R_g \in [-2.3, 0]$ (Fig.~\ref{fig:var_interaction_density}(c),$\varphi_{DE} = 0.4$). Fig.~\ref{fig:var_interaction_density}(d)(e) illustrates the density distribution of the cholesterol headgroup. Membrane thinning with a more attractive $\chi_{AB}N$ also causes a similar redistribution of the cholesterol headgroup. These redistributions occur regardless of whether the cholesterol concentration is $\varphi_{DE} = 0.2$ or $\varphi_{DE} = 0.4$. Similar to the effects on cholesterol chemical potential and phospholipid tilt, $\chi_{AD}N$ and $\chi_{BD}N$ have negligible influence on the density distribution of the bilayer. A detailed density distribution map can be found in Appendix\ref{Appendix:Density}.

\section{Conclusion}
\label{SCE:Conclusion}

In this study, we modeled a single species of phospholipid and cholesterol bilayer as two distinct diblock copolymers. A key feature of our model is the use of separate polymer chains to represent the solvent, phospholipid headgroup, and cholesterol headgroup, while the hydrophobic segments are modeled as rigid rods. Rather than relying on explicit molecular structures, the model captures their physical consequences by encoding interactions within a canonical ensemble and applying self-consistent field theory. Using volume fractions and chain lengths reflective of cholesterol and DPPC geometries, we found that the $C_s$-phase bilayer—with tilted phospholipid tails, resembling the gel phase of DPPC—has a lower free energy than the A-phase bilayer.

Our approach successfully reproduces several well-known effects of cholesterol on bilayers: the condensation effect in saturated PC membranes, increased membrane thickness, and decreased lipid tilt with increasing cholesterol concentrations. The model also captures the experimentally observed linear relationship between cholesterol concentration and its chemical potential in DPPC/cholesterol liposomes. However, it remains unclear whether the model can predict other thermodynamic properties such as phase transitions or heat capacities.

Polymer stretching in our model governs headgroup interactions, but the mapping between polymer deformation and real molecular interactions is not yet fully understood. Due to the large size of the phospholipid headgroup relative to cholesterol, interactions between the phospholipid headgroup and solvent significantly influence membrane behavior, while those involving the cholesterol headgroup have negligible effects. Notably, the minimal influence of phospholipid-cholesterol headgroup interactions contrasts with experimental evidence, which shows that headgroup structure plays a critical role in membrane phase behavior and ligand binding \cite{samsonov2001characterization}. A more detailed model, developed within the same theoretical framework, may help elucidate how headgroup structure governs membrane function and support the design of synthetic biomimetic membranes.

In conclusion, our model provides several advantages: it quantitatively captures cholesterol’s effect on chemical potential and qualitatively reproduces structural trends in lipid bilayers. However, it is currently limited to saturated phospholipid systems. Future work will introduce bent-core molecules to mimic unsaturated lipids. Ultimately, we aim to model cholesterol’s impact on multi-component lipid bilayers and to predict their phase behavior.

%To print the credit authorship contribution details
\printcredits

\section*{Declaration of competing interest}
The authors declare that they have no known competing financial interests or personal relationships that could have appeared to influence the work reported in this paper.

\section*{Acknowledgements}
FS Cohen is supported by the National Institutes of Health (Grant No. 5RO1GM136777).
SX Xu is supported by the National Natural Science Foundation of China (Grant No. 12271492).
YQ Cai is supported by the National Natural Science Foundation of China (Grant No. 12201053).

%\bibliographystyle{cas-model2-names}
%
%% Loading bibliography database
%\bibliography{cas-refs}

\appendix
\section{Appendix}

\subsection{Energy classification}
\label{App:Energy classification}
The free energy includes three parts: the Flory-Huggins internal energy $\F_{FH}$, the orientational interaction $\F_{MS}$, and the entropy $-TS$. These can be defined as \cite{matsen2001standard}
\begin{align}
\F_{FH}
=
      \frac{\tilde\rho}{V}\int_V d\mathbf{r}
\big[\sum\limits_{i \neq j}\chi_{ij}N \phi_i(\mathbf{r})\phi_j(\mathbf{r})
       \big
       ],
\end{align}
\begin{align}
\F_{MS}
=
      \frac{\tilde\rho}{V}\int_V d\mathbf{r}
\big[
              -\frac{\eta N}{2}\S(\mathbf{r}):\S(\mathbf{r})
     \big],
\end{align}
and the entropy:
\begin{align}
-TS
=
      \frac{\tilde\rho}{V}\int_V d\mathbf{r}
\big[&
       - \sum\limits_{i }\omega_i(\mathbf{r})\phi_i(\mathbf{r})
   +    \M(\mathbf{r}):\S(\mathbf{r})
             \big]\notag\\
      &\quad - \log (Q_{A}^{\n_A} Q_{BC}^{\n_{BC}} Q_{DE}^{\n_{DE}}).
\end{align}
The contribution to entropy comes from three parts: the entropy loss arising from the stretching of A-Coil, B-Coil, and D-Coil, $E^{A}, E^{B}$, and $E^{D}$. These contributions can be divided as follows:
\begin{align}
E^{A}
=
      \frac{\tilde\rho}{V}\int_Vd\mathbf{r}
\big[
       - \omega_A(\mathbf{r})\phi_A(\mathbf{r}) \big]
       - \n_A\log (Q_{A}) ,
\end{align}
\begin{align}
E^{B}
=
      \frac{ \tilde\rho}{V}\int_Vd\mathbf{r}
\big[&
       - \omega_B(\mathbf{r})\phi_B(\mathbf{r})
       - \omega_C(\mathbf{r})\phi_C(\mathbf{r})\notag\\
         & + \M(\mathbf{r}):\S_C(\mathbf{r})
        \big]
       - \n_{BC}\log (Q_{BC}).
\end{align}
\begin{align}
E^{D}
=
      \frac{\tilde\rho}{V}\int_Vd\mathbf{r}
\big[&
       - \omega_D(\mathbf{r})\phi_D(\mathbf{r})
       - \omega_E(\mathbf{r})\phi_E(\mathbf{r})\notag\\
       & + \M(\mathbf{r}):\S_E(\mathbf{r})
       \big]
       -\n_{DE}\log(Q_{DE})
\end{align}

\subsection{Computation domain}
\label{App: Computation domain}
We execute the simulation of a self-assembled bilayer on a one-dimensional spatial domain. We denote the membrane area as $S = V/l$ and the computation domain as $[-\frac{l}{2}, \frac{l}{2}]$. In this case, the average over $V$ is simply the average over $[-\frac{l}{2}, \frac{l}{2}]$. For example the first term in \eqref{energy:Hamiltonian1} becomes
\begin{align}
&\frac{1}{V} \int_V\chi_{AB}N\phi_A(\r)\phi_B(\r) d\r \notag\\
&\quad\quad\quad\quad\quad\quad= \frac{1}{l} \int_{-\frac{l}{2}}^{\frac{l}{2}}\chi_{AB}N\phi_A(x)\phi_B(x) dx.
\end{align}
In general, for any reasonable value of $l$, we can solve the SCF equations to obtain the free energy $\frac{NF}{\rho_0V}$. Therefore, we write the energy as a function of $l$, \emph{i.e.,} $F(l)$. As we are considering the self-assembled bilayers, we do not assign any special value to $l$ or the area of the bilayer to $S$.  Instead, we adjust $l$ such that $F(l)$ achieves its minimum at $l^{*}$, which corresponds to the equilibrium states (stable or metastable).  We call $l^{*}$ the optimal computational domain. Unless otherwise stated, the energies in this paper are taken at $l^{*}$.

\subsection{Chemical potential of DE-polymers}
\label{App: CH_DE}

Denoting the free energy as a function of $\n_{DE}$, the definition of $\mu$ reads
\begin{align}
\mu = \frac{\partial \mathcal{F}}{\partial n_{DE}} = \frac{\partial}{\partial n_{DE}}\left(\frac{\rho_0 V F}{N}\right) :=\frac{\partial \mathcal{\tilde{F}}}{\partial \tilde{n}_{DE}}.
\end{align}
In this paper, a bilayer in which the number of DE-polymers is $\n_{DE} = 0$ is the reference state for the chemical potential of DE-polymers, unless otherwise stated. That is, the chemical potential relative to this state is $\Delta  \mu = \mu(\n_{DE}) - \mu(\n_{DE} = 0)$.

\subsection{Lipid tilt}
\label{App: tilt}
The orientational density distribution $\S(x)$ is a second-order tensor. We can estimate a rod's tilt based on the spectral decomposition of the matrix. For example, $\S_C(x)$ can be written as:
\begin{align}
\S_C(x) = \lambda_1 \mathbf{n}_1\mathbf{n}_1 + \lambda_2 \mathbf{n}_2\mathbf{n}_2 + \lambda_3 \mathbf{n}_3\mathbf{n}_3,
\end{align}
where $\lambda_1$, $\lambda_2$, and $\lambda_3$ are the three eigenvalues of $\S_C(x)$, and $\mathbf{n}_1$, $\mathbf{n}_2$, $\mathbf{n}_3$ are the corresponding eigenvectors. The C-phase bilayer corresponds to a liquid crystal in the uniaxial state \cite{cai2017liquid}. Assuming $\lambda_1 > \lambda_2 = \lambda_3$, the expression for $\S_C(x)$ becomes:
\begin{align}
\S_C(x) = \frac{3}{2}s(x)\left(\mathbf{n}_1\mathbf{n}_1 - \frac{I}{3}\right),
\end{align}
where $\mathbf{n}_1$ represents the orientation of the lipid tail in the bilayer. Additionally, the angle between $\mathbf{n}_1$ and the x-axis can be calculated to obtain the average tilt of the lipid. Another way to determine the tilt is by using the thickness ratio between the A-phase and C-phase; however, this approach requires information about the A-phase. Therefore, we use the tilt derived directly from the orientational density distribution $\S(x)$.

\subsection{The density concentration of a bilayer for different $\chi_{AD} N$ and $\chi_{BD} N$ }
\label{Appendix:Density}

\begin{figure}[htbp!]
\centering
\includegraphics[width= 1\linewidth]{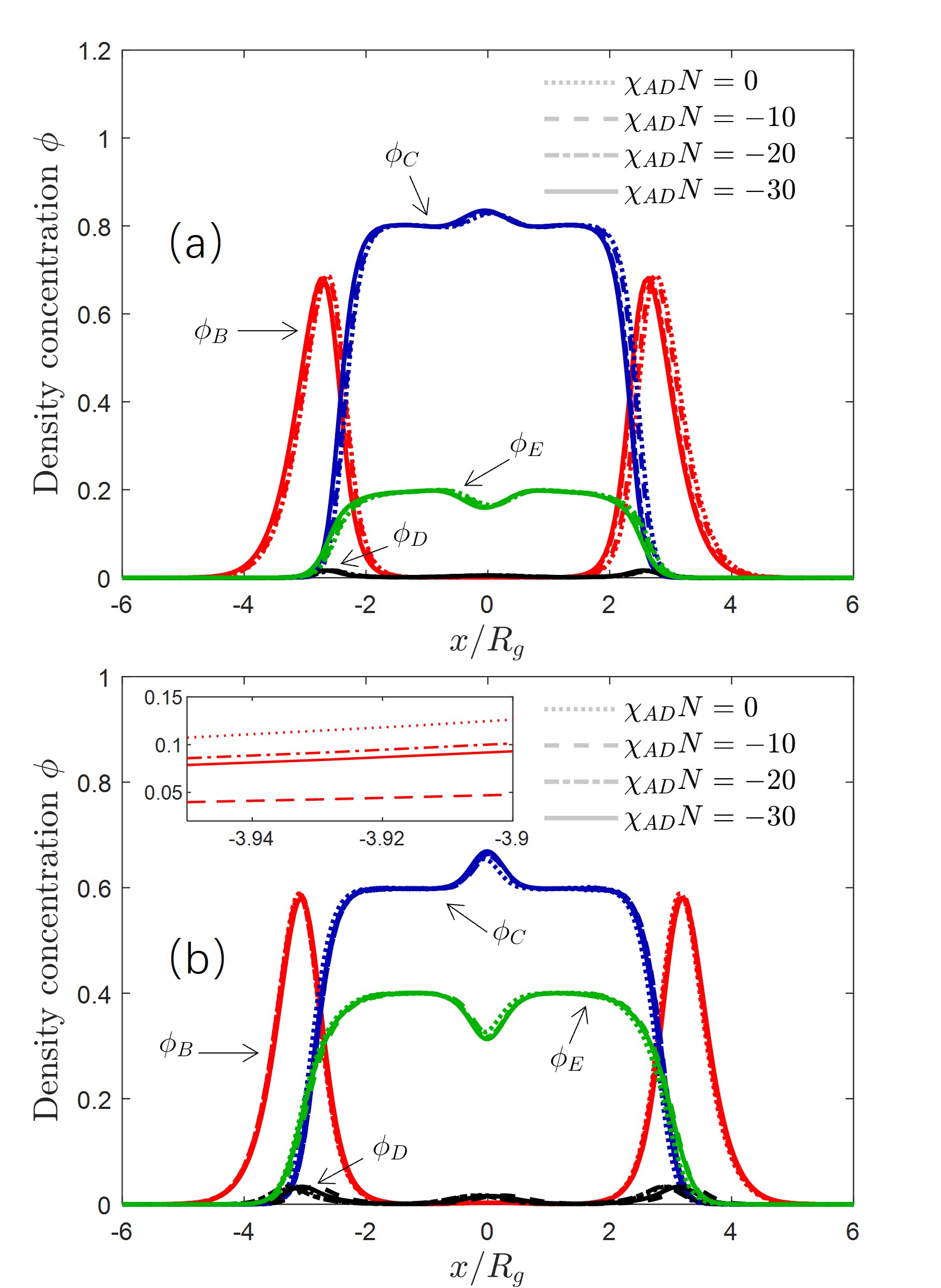}
\caption{The density concentration of the bilayer for different values of $\chi_{AD} = 0$ (dotted curve), $-10$ (dashed curve), $-20$ (dash-dotted curve), $-30$ (solid curve) at cholesterol concentrations of $\varphi_{DE} = 0.2$ (a) and $\varphi_{DE} = 0.4$ (b). The inset in (b) zooms into the density concentration of the phospholipid headgroup at $x/R_g \in[-3.95,-3.9]$. The red, blue, black, and green curves correspond to the phospholipid headgroup, phospholipid tail, cholesterol headgroup, and cholesterol tail, respectively. The near-complete overlap of the dotted, dashed, dash-dotted, and solid curves indicates that variations in$\chi_{AD}$ exert a negligible effect on the density distribution.}
\label{fig:varinteraction_density_Appdi_1}
\end{figure}
\begin{figure}[htbp!]
\centering
\includegraphics[width= 1\linewidth]{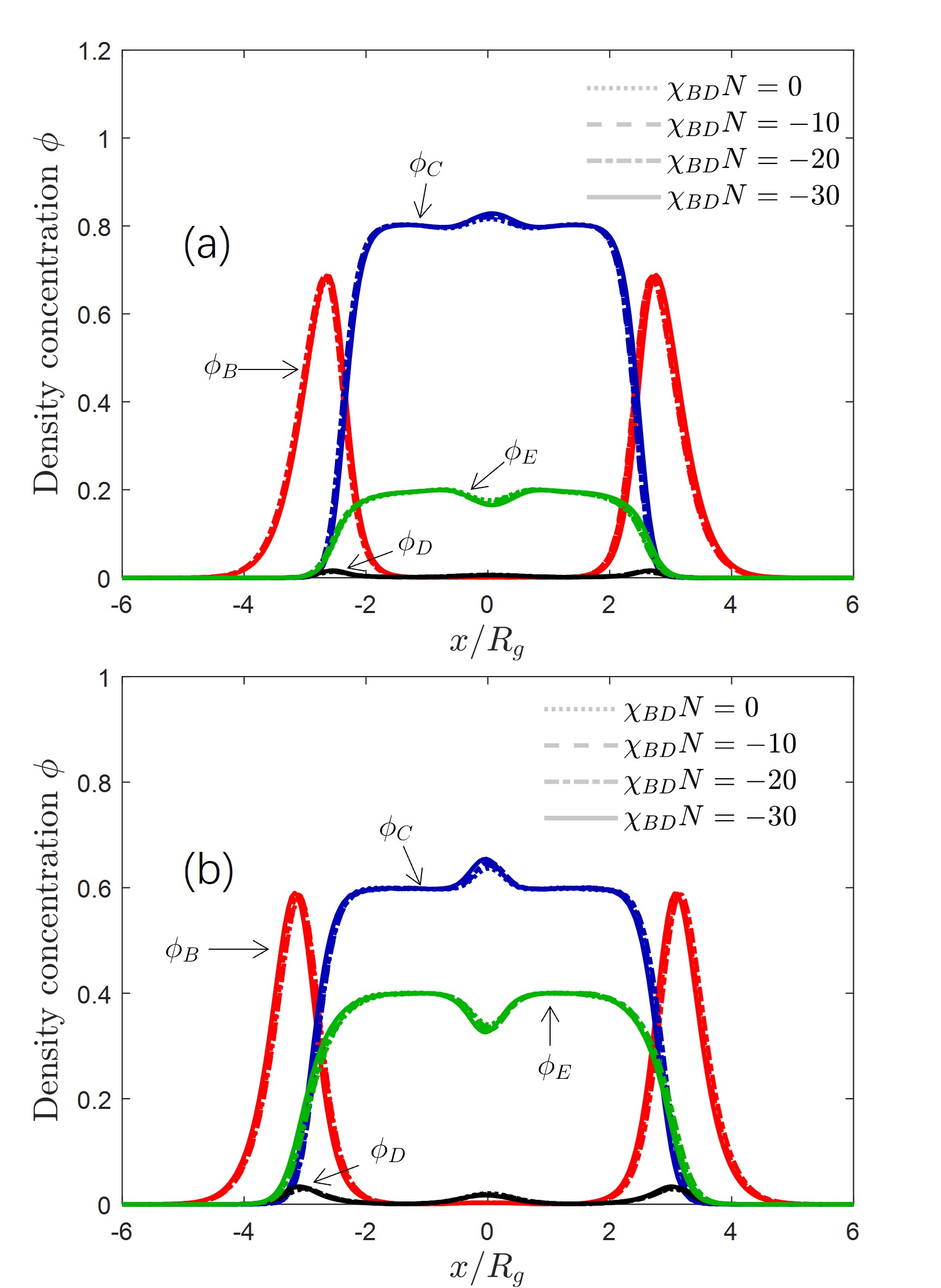}
\caption{The density concentration of the bilayer for different values of $\chi_{BD} = 0$ (dotted curve), $-10$ (dashed curve), $-20$ (dash-dotted curve), $-30$ (solid curve) at cholesterol concentrations of $\varphi_{DE} = 0.2$ (a) and $\varphi_{DE} = 0.4$ (b).
The red, blue, black, and green curves correspond to the phospholipid headgroup, phospholipid tail, cholesterol headgroup, and cholesterol tail, respectively. The close overlap among the dotted, dashed, dash-dotted, and solid curves suggests that variations in$\chi_{BD}$ exert a negligible effect on the density distribution.}
\label{fig:varinteraction_density_Appdi_2}
\end{figure}

The density concentration of the bilayer is shown in Fig.~\ref{fig:varinteraction_density_Appdi_1} for different values of $\chi_{AD}$ at cholesterol concentrations of $\varphi_{DE} = 0.2$ and $\varphi_{DE} = 0.4$. The density concentration of the bilayer is shown in Fig.~\ref{fig:varinteraction_density_Appdi_2} for different values of $\chi_{BD}$ for cholesterol concentrations of $\varphi_{DE} = 0.2$ and $\varphi_{DE} = 0.4$.  The near-complete overlap of the dotted, dashed, dash-dotted, and solid curves indicates that $\chi_{AD}N$ and $\chi_{BD}N$ have minimal impact on the density distributions for a bilayer.

\end{document}